\documentclass[pra, amsfonts, amssymb, amsmath, reprint, showkeys, nofootinbib]{revtex4-1}
\usepackage[english]{babel}
\usepackage[utf8]{inputenc}
\usepackage[colorinlistoftodos, color=green!40, prependcaption]{todonotes}
\usepackage{amsthm}
\usepackage{physics}
\usepackage{xcolor}
\usepackage{graphicx}
\usepackage[left=23mm,right=13mm,top=35mm,columnsep=15pt]{geometry} 
\usepackage{adjustbox}
\usepackage{placeins}
\usepackage[T1]{fontenc}
\usepackage{lipsum}
\usepackage{csquotes}
\usepackage{mathbbol}
\usepackage[pdftex, pdftitle={Article}, pdfauthor={Author}]{hyperref} 
\begin{document}
\title{Computational Phase Transition Signature in Gibbs Sampling}

\author{H.~Philathong}
    \email[e-mail:]{hariphan@skoltech.ru}
    \homepage{http://quantum.skoltech.ru}
    \affiliation{Deep Quantum Laboratory, Skolkovo Institute of Science and Technology, 3 Nobel Street, Moscow, Russia 121205}
\author{V.~Akshay}
    \affiliation{Deep Quantum Laboratory, Skolkovo Institute of Science and Technology, 3 Nobel Street, Moscow, Russia 121205}
\author{I.~Zacharov}
    \affiliation{Deep Quantum Laboratory, Skolkovo Institute of Science and Technology, 3 Nobel Street, Moscow, Russia 121205}
\author{J.D.~Biamonte}
    \affiliation{Deep Quantum Laboratory, Skolkovo Institute of Science and Technology, 3 Nobel Street, Moscow, Russia 121205}

\date{\today} 

\begin{abstract}
Gibbs sampling is fundamental to a wide range of computer algorithms.  Such algorithms are set to be replaced by physics based processors---be it quantum or stochastic annealing devices---which embed problem instances and evolve a physical system into an ensemble to recover a probability distribution.   At a critical constraint to variable ratio, decision problems---such as propositional satisfiability---appear to statistically exhibit an abrupt transition in required computational resources.  This so called, algorithmic or computational phase transition signature, has yet-to-be observed in contemporary physics based processors.  We found that the computational phase transition admits a signature in Gibbs' distributions and hence we predict and prescribe the physical observation of this effect.  We simulate such an experiment, that when realized experimentally, we believe would represent a milestone in the physical theory of computation.
\end{abstract}

\maketitle


As information is necessarily represented in physical media, the processing, storage and manipulation of information is governed by the laws of physics. Indeed, the theory of computation is intertwined with the laws governing physical processes \cite{deutsch1985quantum}.  Many physical systems and physical processes can be made to represent, and solve computational problem instances.  Viewed another way, many variants of naturally occurring processes (such as protein folding) have been shown to represent computationally significant problems, such as {\bf NP}-hard optimization problems.  But how long does it take for a physical process to solve problem instances?  How can difficult problem instances be generated? 

The physical Church-Turning thesis \cite{church1936unsolvable, turing1937computable} asserts that a universal classical computer can simulate any physical process and vise versa (outside of quantum mechanical processes).  It does not propose the algorithm, yet asserts its existence.  One might wrongly suspect that undecidable problems can be embedded into physical systems: attempts at this fail, i.e.~due to instabilities. What about the {\bf P} vs.~{\bf NP} problem?  If no physical process existed to solve {\bf NP}-complete problems in polynomial time, then by the physical Church-Turning thesis, no algorithm would exist either.  Hence if the laws of physics ruled out such a scenario, this would imply that {\bf P $\neq$ \bf NP}.  

This distinction between polynomial and exponential resources is a course gaining that computational complexity theory is based around. We do not know if a physical process can be made to solve {\bf NP}-complete problems in polynomial time or not. However, it is asserted that computational phase transitions are a  feature of {\bf NP}-complete problems---although specifics of the transition have yet to be formulated (proven) rigorously. We will turn to the theory of computational phase transitions to understand how physics responds to changes in the complexity landscape across the algorithmic phase transition.


This algorithmic phase transition occurs where randomly generated problem instances are thought to be difficult \cite{crawford1996experimental, friedgut1999sharp, selman1996critical}.  It is observed by the fact that computer algorithms experience a slowdown around this transition point.  For example, let us consider the familiar problems of $2$- (and $3$)-satisfiability (detailed in the next section: the phase transition provably exists for $2$-SAT and is only known to be inside a window for $3$-SAT).  If we let the number of variables be $N$ and uniformly generate $M$ random clauses over these $N$ variables, computer algorithms appear to slow down at a certain clause to variable ratio (clause density $\alpha = M/N$). 

What about physical systems that bootstrap physics to naturally solve problems instances? Instances of these problems can be embedded in the lowest energy configuration of physical models \cite{lucas2014ising, biamonte2008nonperturbative, whitfield2012ground}. Hence, building such a physical system and cooling (annealing) this system can enable a process which solves such problems \cite{kirkpatrick1983optimization, utsunomiya2011mapping, inagaki2016coherent, pierangeli2019large, marandi2014network, nixon2013observing, berloff2017realizing, dung2017variable, kalinin2018global}.  For example, a system settling into its low-energy configuration can be programmed such that this low energy configuration represents the solution to $2$ (and $3$)-SAT instances.  

We found that the algorithmic phase transition has a statistical signature in Gibbs' states of problem Hamiltonians generated randomly across the algorithmic phase transition.  This was confirmed by exact calculations of 26 binary units (spins) on a mid-scale supercomputer.  Physical observation of the effect is hence within reach of near term and possibly even existing physical computing hardware: such as Ising machines \cite{inagaki2016coherent, pierangeli2019large}, annealers \cite{kirkpatrick1983optimization} and quantum enhanced annealers \cite{ johnson2011quantum, barends2016digitized, harris2010experimental,harris2018phase,king2018observation}.

\paragraph*{Satisfiability Phase Transition Signature.}\label{sec:CPT}
The Boolean satisfiability (SAT) problem is a decision problem determining satisfiability of Boolean formula. If the formula is evaluated to be TRUE for some assignments of its Boolean variables, the formula is called {\it satisfiable}. If there is no such assignment, the formula is {\it unstaifiable}. Deciding satisfiability of a formula in conjunctive normal form (CNF) where each clause is limited to at most $k$ literals is {\bf NP}-complete for $k \geq 3$. It is known that every $k$-SAT problem can be Karp-reduced to the {\bf NP}-complete $3$-SAT problem \cite{cook1971complexity}. We consider optimization of an underlying decision problem.  Hence we seek to determine the maximum number of clauses that can be satisfiable over all assignments of input variables. This problem is called MAX $k$-SAT which is a canonical {\bf NP}-hard optimisation problem.

We will let a $k$-SAT instance consist of $M$ clauses in $k$ conjunctive normal form over $N$ Boolean variables. The clause density of an instance is defined by the simple faction $\alpha=M/N$. For example, consider the $3$-SAT problem instance
\begin{equation}\label{sat_ex}
(x_{1} \lor \neg x_{2} \lor x_{3})\land  (x_{1} \lor x_{4} \lor \neg x_{5}). 
\end{equation}
The instance \eqref{sat_ex} has $5$ variables, $2$ clauses ($\alpha=2/5$) and is satisfiable, as $x_1=1,\,x_2=1,\,x_3=0,\,x_4=0,\,x_5=0$.  $k$-SAT clauses are randomly generated to form random instances by uniformly selecting unique $k$-tupels from the union of a variable set (cardinality $N>k$) and its element wise negation.  

\begin{figure*}[]

\minipage{0.5\textwidth}

  \includegraphics[width=\linewidth,height=5cm]{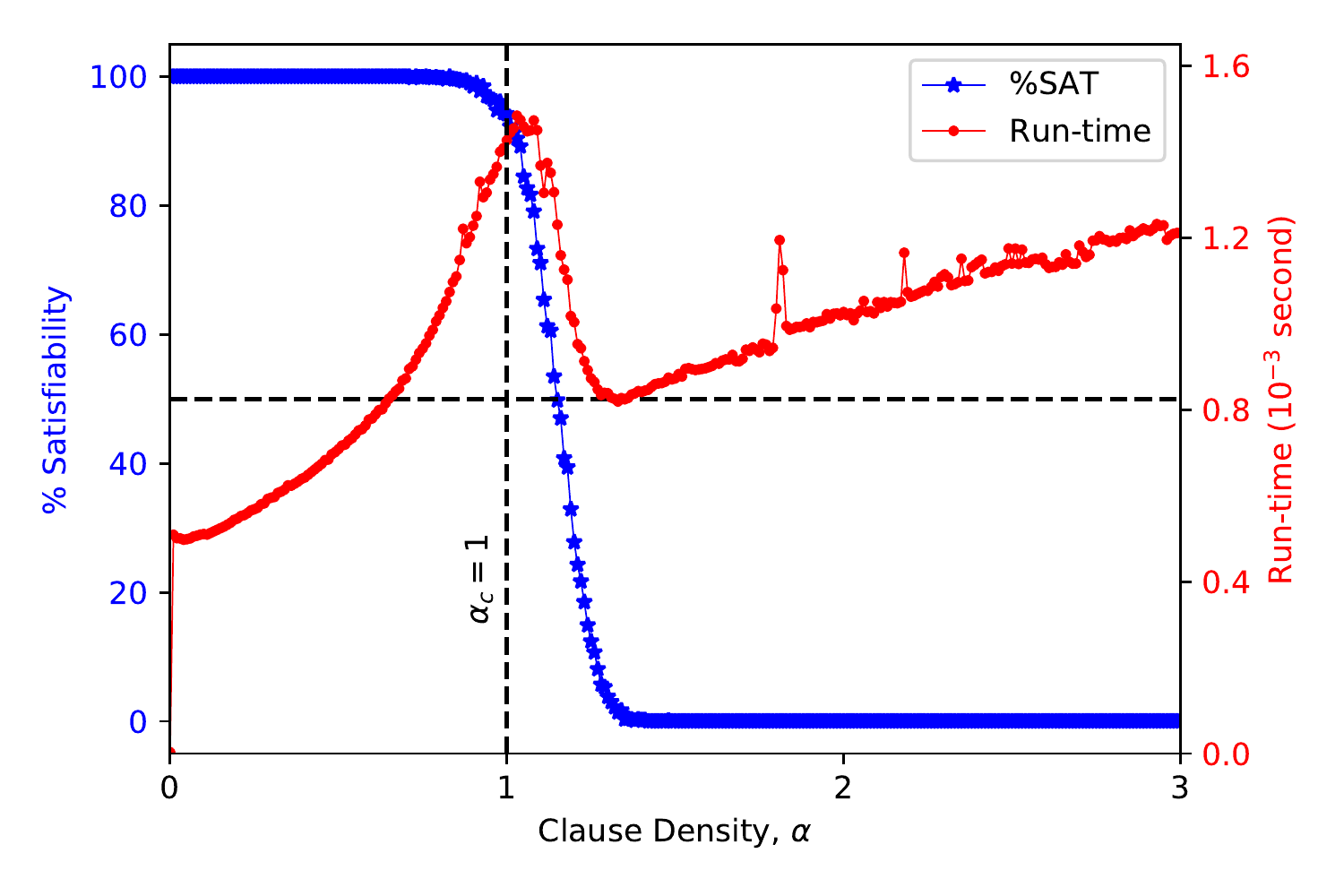}
\endminipage
\minipage{0.5\textwidth}
  \includegraphics[width=\linewidth,height=5cm]{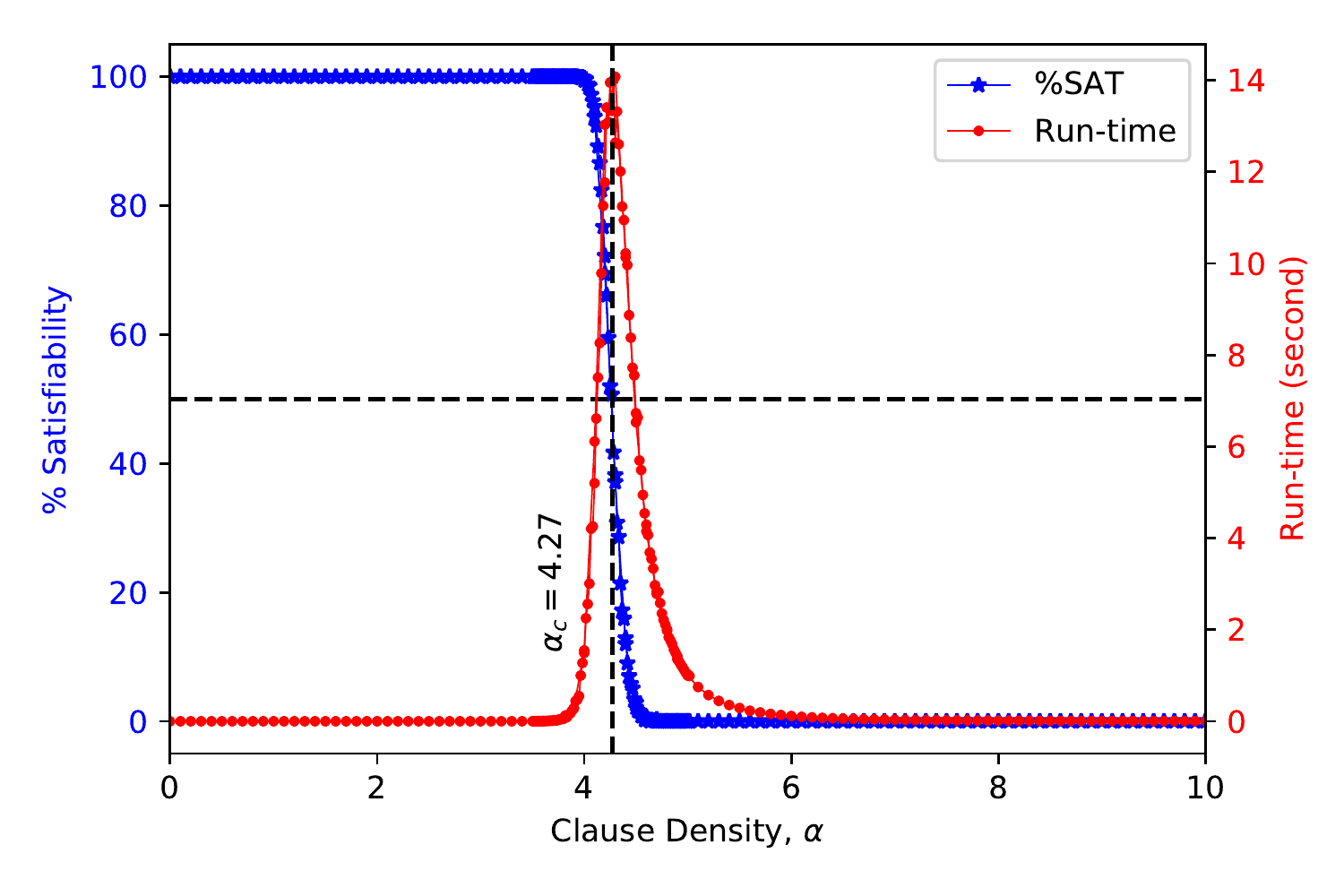}
\endminipage

\minipage{0.5\textwidth}
  \includegraphics[width=\linewidth,height=5cm]{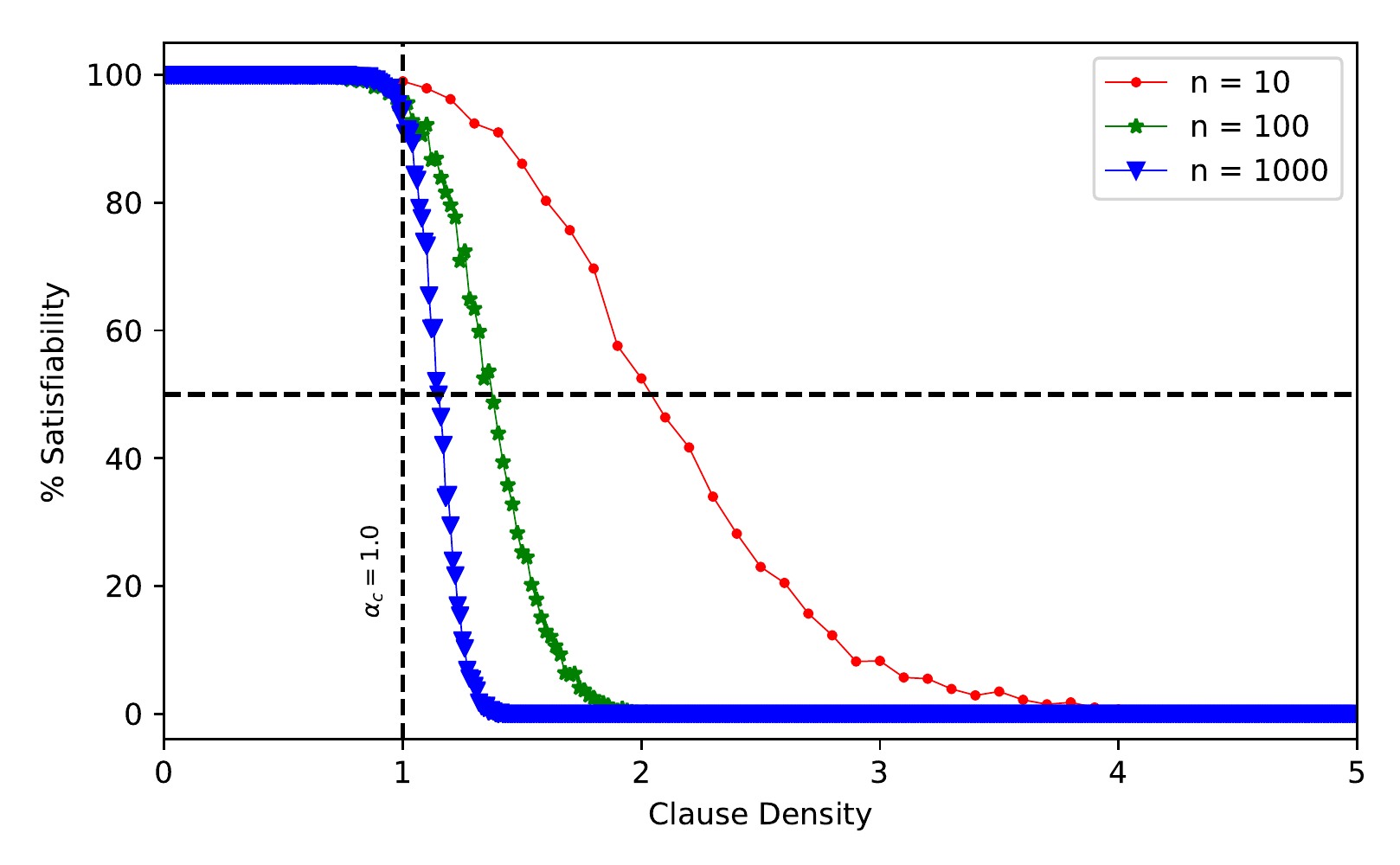}
\endminipage
\minipage{0.5\textwidth}
  \includegraphics[width=\linewidth,height=5cm]{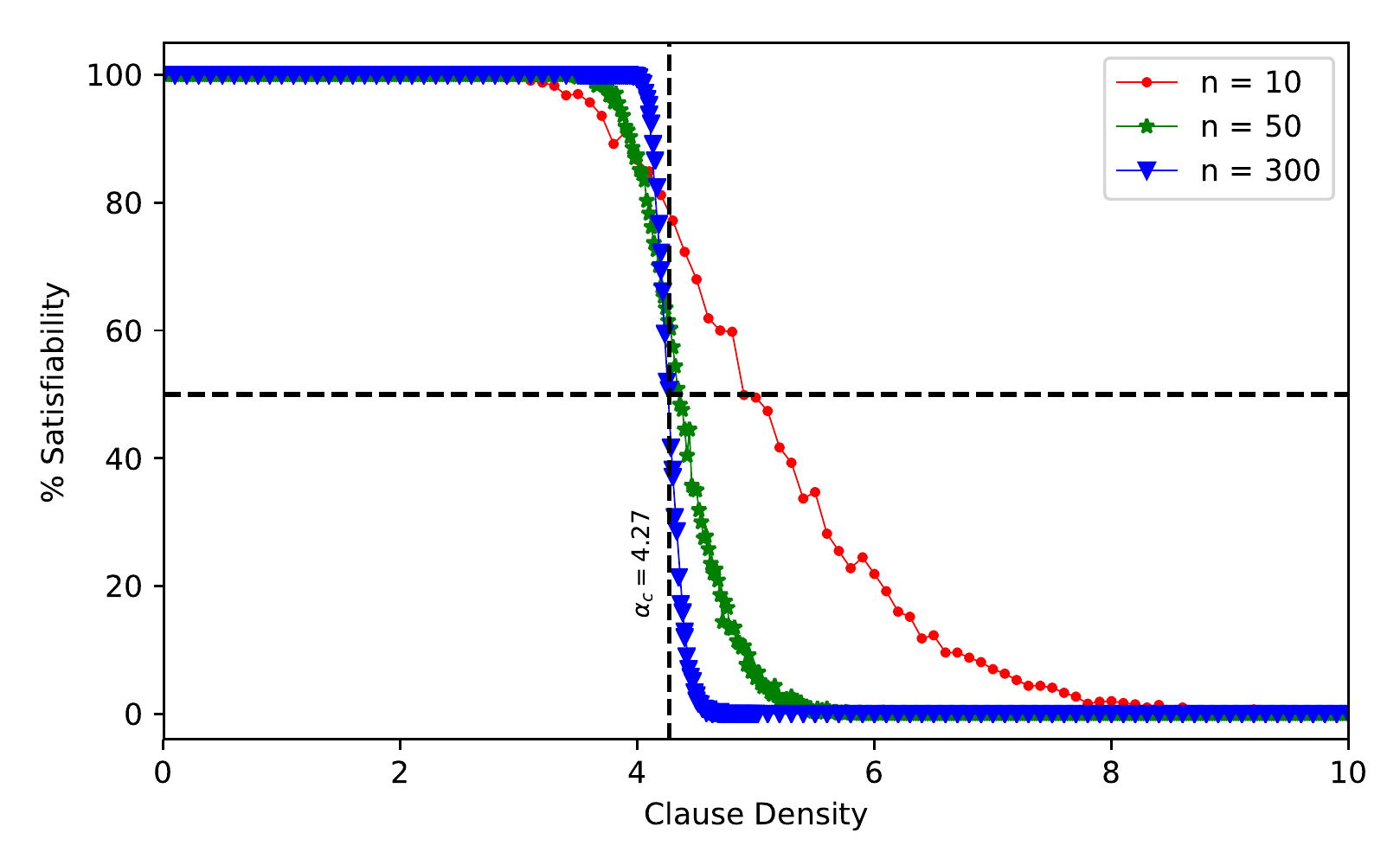}
\endminipage

\minipage{0.5\textwidth}
  \includegraphics[width=\linewidth,height=5cm]{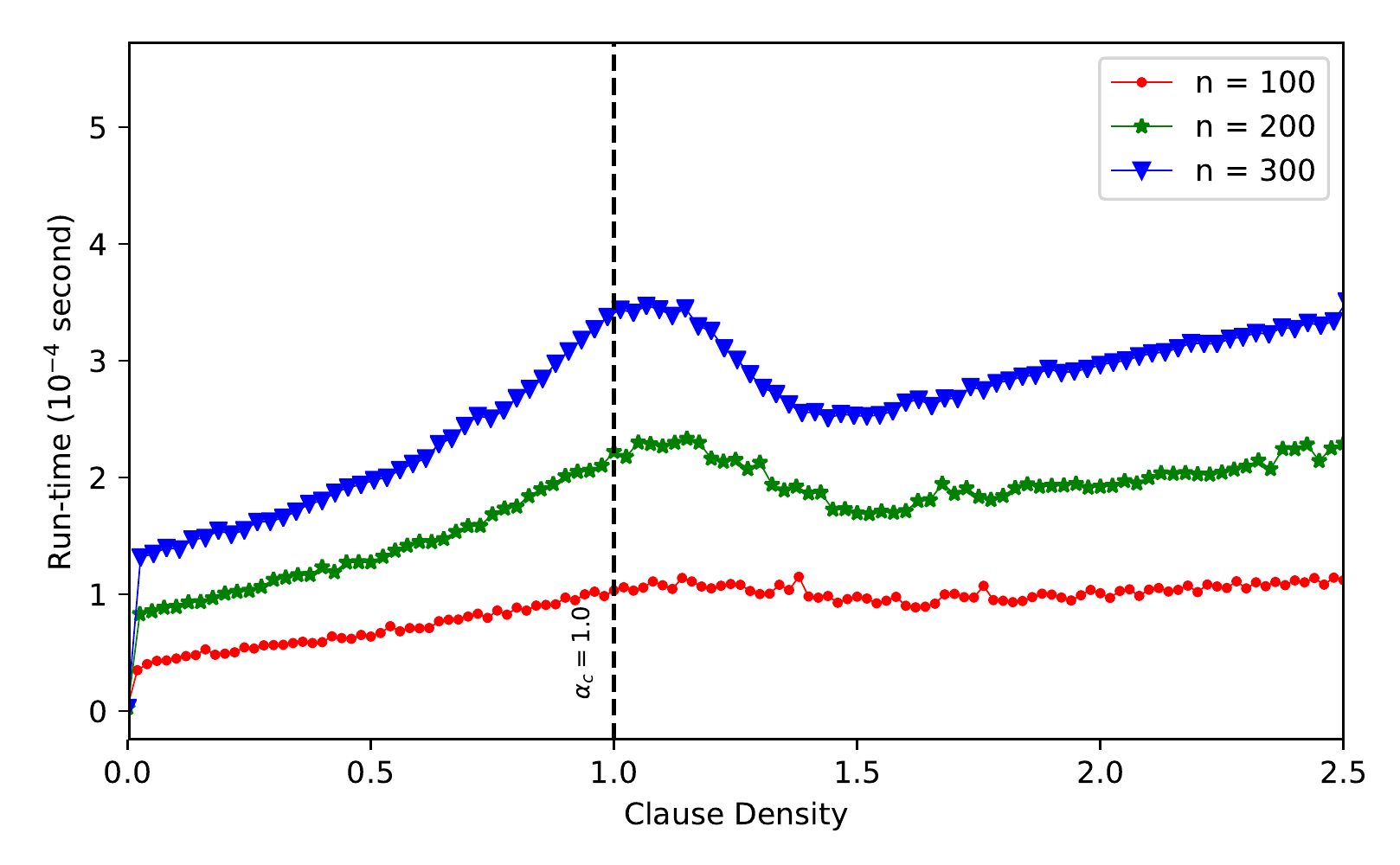}
\endminipage
\minipage{0.5\textwidth}
  \includegraphics[width=\linewidth,height=5cm]{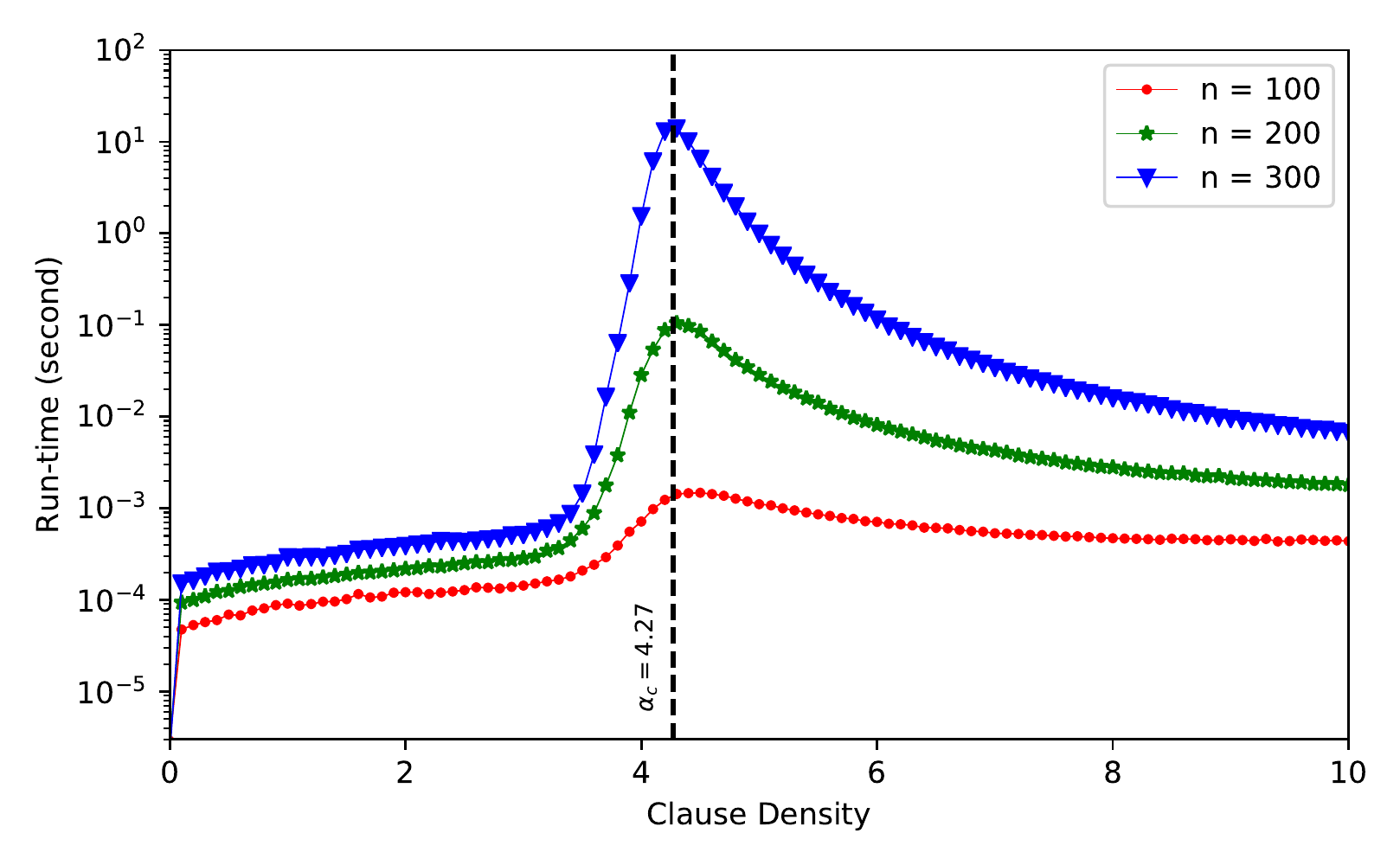}
\endminipage
\caption{(Top): Percent of satisfiable instances (left axis) and run-time (right axis)) versus clause density. Varying clause density $\alpha$, we randomly generated $1,000$ $2$-SAT instances  with $1,000$ variables (left) and $3$-SAT instances with $300$ variables (right). Here $\alpha_c = 1$ and $\alpha_c = 4.27$ for $2$-SAT and $3$-SAT respectively. (Middle): Percent of satisfiability of 2-SAT instances (left) and 3-SAT instances (right) across the algorithmic phase transition signature, comparing total variable count. (Bottom): Run-time of 2-SAT instances (left) and 3-SAT instances (right) across the algorithmic phase transition signature, comparing total variable count.
}
\label{Fig:CPT}
\end{figure*}

For random $3$-SAT instances, an abrupt transition in the probability of satisfiability has been numerically observed (though not formally proven) to be around clause density $\alpha_{c}\sim 4.27$ \cite{crawford1996experimental,friedgut1999sharp,selman1996critical,selman1996generating}. Most instances on the left (right) of the transition are satisfiable (unsatisfiable). It is statistically observed that computational time of SAT solvers peak at the same point (see Fig.~\ref{Fig:CPT} (top right)). These two features indeed show the signature of the so called computational (or algorithmic) phase transition. The requirement of increasing computational resources to solve instances at and around this transition point suggests that hard instances for the $3$-SAT problems are concentrated around a critical density.

The computational phase transition signature has better theoretical footing when considering the $2$-SAT problem, which can be solved in polynomial time \cite{krom1967decision}. Both theoretical and numerical results \cite{bollobas2001scaling,chvatal1992mick,goerdt1996threshold} firmly position the transition at clause density $\alpha_{c}= 1$ (see Fig.~\ref{Fig:CPT} (top left)).  The further advantage is that additional slack bits are not required to embed 2-SAT into the Ising model.  In other words, 2-SAT instances can be embedded directly into 2-body generalized Ising Hamiltonians (see Appendix). 

We randomly generate $1,000$ instances of 2 (and 3)-SAT then calculate the percent of satisfiability (see Fig.~\ref{Fig:CPT} (middle)) and run-time (see  Fig.~\ref{Fig:CPT} (bottom)) using a SAT solver, PycoSat package in Python 3 \cite{biere2008picosat}. The sharp transition at the critical point $\alpha_c$ can be observed in infinite limit of a ratio. In practice, finite sized effects broaden the transition region.

For 2-SAT, the theoretical results \cite{bollobas2001scaling} establish a finite-size scaling window for the transition region. For $0 < \delta < 1$, this window is defined by $W(N,\delta)=(\alpha_{-}(N,\delta),\alpha_{+}(N,\delta))$, where $\alpha_{-}(N,\delta)=\sup(\alpha)$ such that the probability of satisfiablity is greater than $1-\delta$, similarly, $\alpha_{+}(N,\delta)=\inf(\alpha)$ such that the probability of satisfiablity is less than $\delta$. As the width of the window tends to zero, $\alpha_{\pm}(N,\delta) \longrightarrow 1$, as $N \longrightarrow \infty$ for all $\delta$. 

For 3-SAT, it is still unknown where the exact transition point is. However, upper and lower bounds have been proven. The best lower bound at this time appears to be $\alpha=3.52$ \cite{hajiaghayi2003satisfiability} and the best upper bound is $\alpha=4.453$ \cite{maneva2008satisfiability}.

In Fig.~\ref{Fig:CPT} (middle and bottom), we can see that, as the number of variables increase, the transition  sharpens and becomes closer to clause density 1 for 2-SAT (left) and $4.27$ for 3-SAT (right).


\paragraph*{Ground State Occupancy of Thermal States.}\label{sec:thermal} 
Let us develop a technique to witness the computational phase transition in physical systems. We simulate a physical systems that encodes solution(s) of 2 (and 3)-SAT instances into the ground state space of their corresponding Hamiltonians.  This is done by considering the corresponding Gibbs state. We explore the difficulty to sample correct solutions of the problems from the thermal distribution at a finite temperature as a function of clause density. This technique can be formulated as follows.

Let $\ket{i}$ denote (possibly degenerate) lowest eigenstates of $\mathcal{H}$.  We label these possibly degenerate states by letting $i$ range from $1$ up to $d$ and call $\lambda_\text{min}$ the lowest eigenvalue of $\mathcal{H}$
\begin{equation}\label{eqn:lambdamin}
\forall i \in\{1, ..., d\}, ~\min_{\psi} \bra{\psi}\mathcal{H}\ket{\psi}=\bra{i}\mathcal{H}\ket{i}=\lambda_\text{min}.
\end{equation}
Physical systems at thermal equilibrium are said to be ideally described by a Gibbs state, 
\begin{equation}
\rho_\beta = \frac{e^{-\beta \mathcal{H}}}{\mathcal{Z}},
\end{equation}
where $\mathcal{H}$ is the system Hamiltonian (in our case 2 (and 3)-SAT Hamiltonian), $\beta$ is an inverse temperature where the partition function is 
\begin{equation}
\mathcal{Z} = \text{tr}\{e^{-\beta \mathcal{H}}\}.
\end{equation}

Occupancy in the low-energy subspace for a system at equilibrium for fixed finite inverse temperature $\beta$ becomes
\begin{equation}\label{eqn:plowest}
p\left(\lambda_\text{min}, \beta\right) = \frac{1}{\mathcal{Z}}\sum_{i=1}^d\bra{i}e^{-\beta \mathcal{H}}\ket{i}=\frac{d}{\mathcal{Z}}e^{-\beta\lambda_\text{min}}.
\end{equation}

In the zero temperature limit we have 
\begin{equation}
\lim_{\beta \rightarrow \infty}p\left(\lambda_\text{min}, \beta\right)  = \lim_{\beta \rightarrow \infty}\frac{d}{d+\sum\limits_{\lambda_j>\lambda_\text{min}}e^{-\beta(\lambda_j-\lambda_\text{min})}}=1.
\end{equation}

\begin{figure*}[ht!]

\minipage{0.5\textwidth}

  \includegraphics[width=\linewidth,height=5cm]{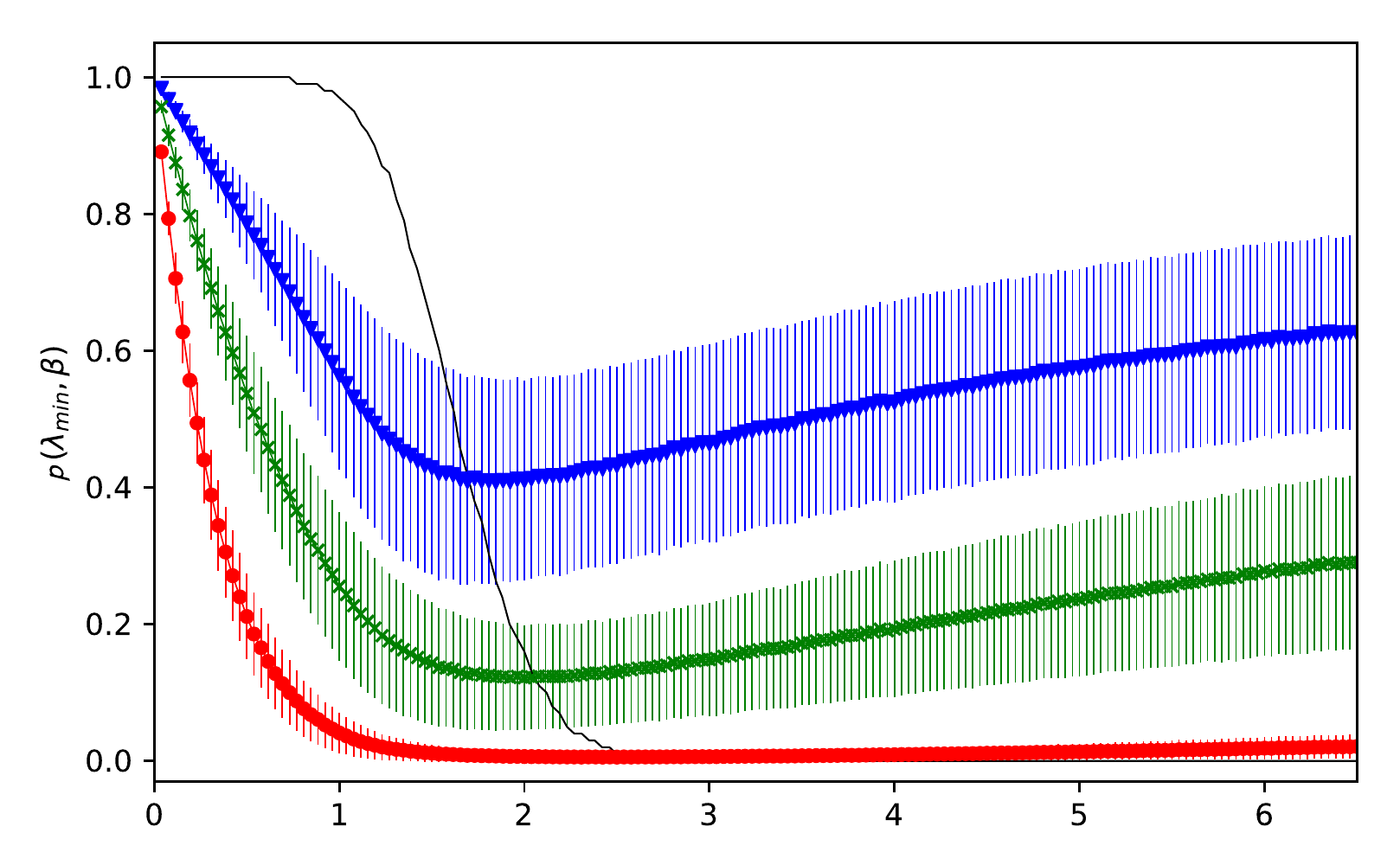}
\endminipage
\minipage{0.5\textwidth}
  \includegraphics[width=\linewidth,height=5cm]{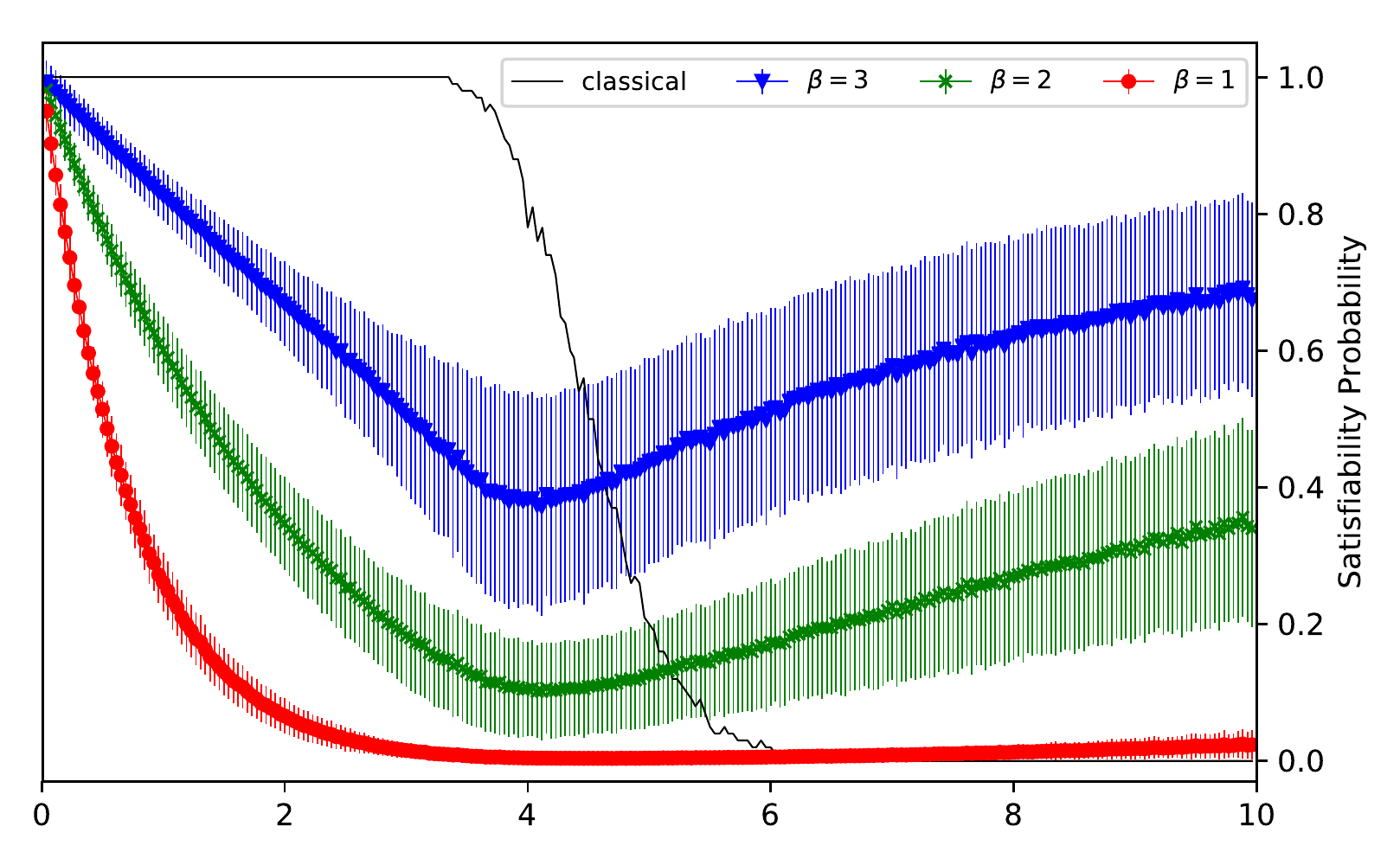}
\endminipage

\minipage{0.5\textwidth}
  \includegraphics[width=\linewidth,height=5cm]{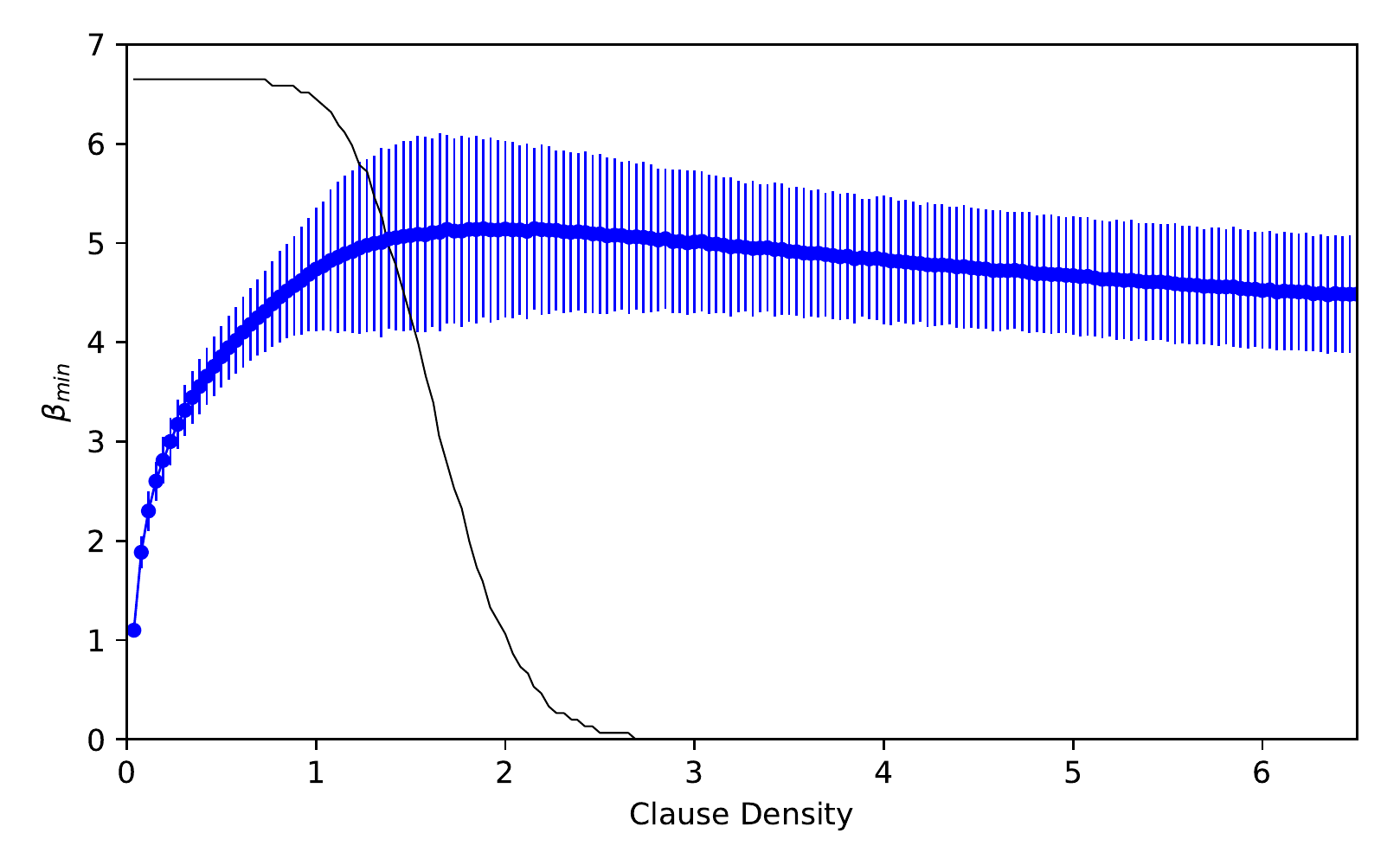}
\endminipage
\minipage{0.5\textwidth}
  \includegraphics[width=\linewidth,height=5.1cm]{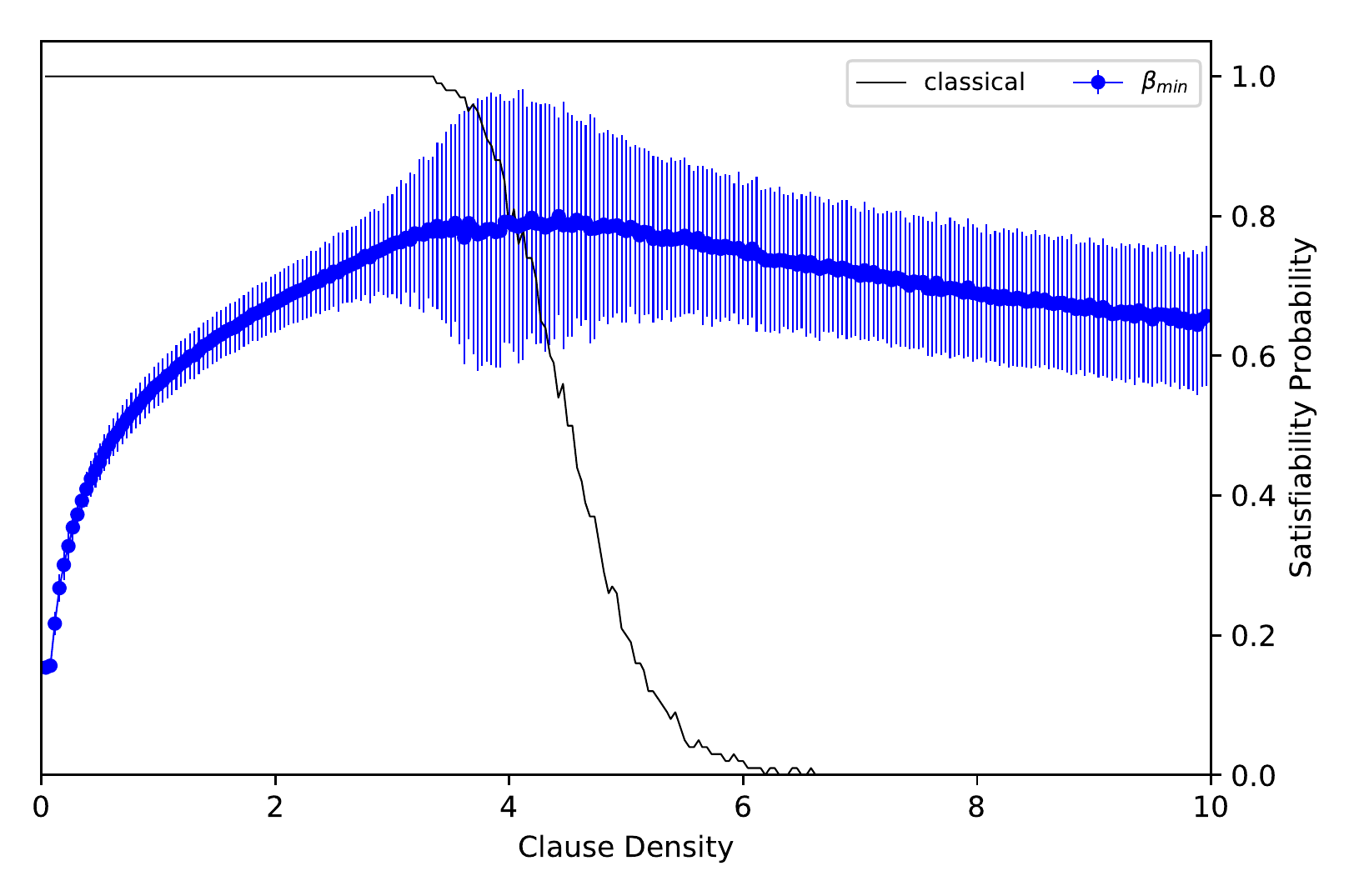}
\endminipage

\caption{(Top): Occupancy of the thermal ground state corresponding to Hamiltonians embedding 2-SAT instances (left) and 3-SAT instances (right) across the algorithmic phase transition for 26 spins with $\beta = 1, 2, 3$. The vertical bars in the plots indicate the standard deviation. (Bottom): Minimum values of $\beta$ (inverse temperature) of $2$-SAT (left) and $3$-SAT (right) such that occupancy of ground state, $p\left(\lambda_\text{min}, \beta\right)$, is greater than $0.9$ across the algorithmic phase transition for 26 spins.
}
\label{Fig:thermal}
\end{figure*}

The quantity $p\left(\lambda_\text{min},\beta \right)$ in Eq.~\eqref{eqn:plowest} provides our means to access the difficulty of sampling the right solution from a thermal state. $p\left(\lambda_\text{min}, \beta\right)$ as a function of clause density for different number of variables and $\beta$ is illustrated in Fig.~\ref{Fig:thermal} (top). 
For each clause density, we randomly generated $1,000$ $2(3)$-SAT instances from $26$ variables.  We then applied the embedding procedure from appendix \ref{appx:embedding} and calculated $p\left(\lambda_\text{min}, \beta\right)$. 


Surprisingly---on average---the lowest values of $p\left(\lambda_\text{min}, \beta\right)$ are around densities $1.0$ to $2.0$ for $2$-SAT and $3.5$ to $4.5$ for $3$-SAT which means that it is difficult to ascertain the optimal solution when sampling around the critical point. The most difficult instances appear near the point where $50\%$ of the formulas are satisfiable.

On average, what inverse temperature is required to ensure that the occupancy of the ground thermal state, $p\left(\lambda_\text{min}, \beta\right)$, is greater than 0.9? We plot the value of $\beta$ (inverse temperature) satisfying the above condition versus clause density (see Fig.~\ref{Fig:thermal} (bottom)) for both $2$-SAT and $3$-SAT.

In Fig.~\ref{Fig:thermal} (bottom), the peaks of the plots are around critical densities for $2$-SAT and $3$-SAT which appear around the same critical points as for standard algorithms. For physical systems preparing Gibbs distributions, this informs us of the significance that temperature plays obtain above 90 percent ground state occupancy. The hardest instances are around the critical point (the peaks) which require lower temperatures for sufficient ground state occupancy. 

\paragraph*{Conclusion.}
We predict that the algorithmic phase transition signature is observable in thermal Gibbs' states, which has implications for contemporary physical computing systems.  The computational phase transition signature has strong theoretical footing when considering the $2$-SAT problem---a problem which offers a direct embedding into 2-body generalized Ising Hamiltonians.  The 3-SAT problem on the other hand requires additional slack bits to create the 3-body projectors to embed the problem.

For the purpose of comparison, we randomly generate both $2$-SAT ($3$-SAT) instances and embed these instances in the generalized Ising Hamiltonian such that the ground-eigenspace is spanned by solutions to the problem instance. Then we calculate occupancy of ground state of thermal states across the algorithmic phase transition signature. We found that the occupancy of ground state(s) decreases around the phase transition point for both $2$-SAT and $3$-SAT (see Fig.~\ref{Fig:thermal} (top)). This finding indicates a difficulty in sampling solutions of the $2$-SAT and $3$-SAT problems concentrated around a critical point.  Moreover one infers that temperature of the Gibbs state is the fleeting computational resource.  Indeed, one argues that for any fixed inverse temperature $\beta$, there exists problem instances that would require significant sampling time to recover the ground-state. 

Our prediction connects the computational phase transition signature, which exhibits an importantly unsolvable feature in the theory of computational complexity, with physical processes that naturally reveal this feature in observable quantities. Since these SAT instances can be directly embedded onto the generalize Ising Hamiltonian, and recently Ising devices have been built with increasing programability, the computational phase transition signature has the potential to be physically observed.

\paragraph*{Numerical Methods.} The large calculations were performed on the Skoltech's supercomputer ``Zhores'' \cite{zach2019}. We thank the CDISE-HPC team.  The code is written in C with nested OpenMP parallelism. The Hamiltonian is expanded in the shared memory of each node and we compute in parallel several clauses. Multiple nodes run the procedure concurrently with different random sequence to sample the statistics with MPI collectives. This implementation can use all the operational memory of each server. In practice we have compared results for up to 34 variables and found no qualitative differences that would affect our conclusions.  
In Fig.~\ref{Fig:thermal} (top) we illustrate results for $N=26$; the program run on 8 nodes to collect statistics for around 13 hours. We also used the SAT solver, PycoSat in Python 3 \cite{biere2008picosat}.

\onecolumngrid

\appendix*
\section{Ising Spin Embedding}\label{appx:embedding}

Finding ground energy and ground state configurations of physical systems such as spin glasses is {\bf NP}-hard \cite{barahona1982computational}. 2-SAT and 3-SAT instances can be directly mapped onto the Hamiltonian minimization problem, see e.g.~\cite{lucas2014ising,biamonte2008nonperturbative}.  The standard procedure is as follows.  

We convert logical bits $\{0,1\}$ to spins
\begin{equation}
\text{Logical}\,\ 0 \mapsto \ket{0},\quad \text{Logical}\,\ 1 \mapsto \ket{1}.
\end{equation}
The penalty Hamiltonian $\mathcal{H}_{SAT}$ is constructed by real linear extension of $\{P_0, P_1, \mathbb{1}\}$ by means of the following invertible mapping between Boolean variables and projectors,
\begin{equation}
x_j \longrightarrow P_j^0, \qquad \neg x_j \longrightarrow P_j^1,
\end{equation}
and
\begin{equation}
\wedge \longrightarrow +, \qquad \vee \longrightarrow \otimes,
\end{equation}
where $P_j^{1}=\ketbra{1}{1}$ and $P_j^{0}=\ketbra{0}{0}$ acting on the $j^{th}$ spin.

By this construction, clauses in a SAT instance are mapped onto projectors $P^{\alpha \beta ...\gamma }_{i j...k}$. Hence, the Hamiltonian $\mathcal{H}_{SAT}$ can be constructed by summing over all the clauses in an instance,
\begin{equation}\label{eq:sathamiltonian}
    \mathcal{H}_{SAT}=\sum_{l} \mathcal{C}_{l} \lbrace P_{i j...k}^{\alpha \beta...\gamma } \rbrace,
\end{equation}
where $\mathcal{C}_{l}$ assigns the value of $\alpha, \beta, ..., \gamma \in \{0, 1\}$ corresponding to the $l^{th}$ clause. The ground state space is spanned by solutions of the SAT problem, and the ground energy is equal to minimal number of unsatisfiable clauses.

Substituting the projectors $P^{\alpha}_j=\frac{1}{2}(\mathbb{1} + (-1)^{\alpha} \sigma^{z}_{j})$ in Eq.~\eqref{eq:sathamiltonian},
where $\sigma^{z}$ is the Pauli matrix along the quantization axis, the Hamiltonian $\mathcal{H}_{SAT}$ can be written in the form of the generalized Ising Hamiltonian with k-body interaction. In case of MAX 2-SAT, the Hamiltonian $\mathcal{H}_{2SAT}$ is a sum over projectors onto $2$-SAT clauses 
\begin{eqnarray}
    \mathcal{H}_{2SAT} &&= \sum_l \mathcal{C}_l \lbrace P^{\alpha}_i \otimes P^{\beta}_j\rbrace
    \\
    &&= \frac{1}{4} \left( \sum_{i} \mathbb{1} + \sum_{i}h_{i} \sigma^{z}_{i}   +  \sum_{ij}\mathcal{J}_{ij}\sigma^{z}_{i}\sigma^{z}_{j}\right)
\end{eqnarray}

The coefficient $h_{i}$ indicates the local field of the $i^{th}$ spin and the couplings $\mathcal{J}_{ij}$ encode the 2-body interaction between the $i^{th}$ spin and the $j^{th}$ spin.

This method provides a way to physically realize SAT instances as spin Hamiltonians. If such a physical system is built, cooling the system into its ground state is equivalent to solving the SAT problems.

\end{document}